\documentclass[preprint,aps,amsmath,byrevtex,pra,titlepage,
showpacs,
nofootinbib]{revtex4-1}

\usepackage{amssymb}
\usepackage{bm}

\newcommand{\beq}{\begin{equation}}
\newcommand{\eeq}{\end{equation}}
\newcommand{\eq}[1]{Eq.~(\ref{#1})}

\begin{document}

\title {Weak Interaction Contributions in Light Muonic Atoms}
\author {Michael I. Eides}
\altaffiliation[Also at ]{the Petersburg Nuclear Physics Institute,
Gatchina, St.Petersburg 188300, Russia}
\email[Email address: ]{eides@pa.uky.edu, eides@thd.pnpi.spb.ru}
\affiliation{Department of Physics and Astronomy,
University of Kentucky, Lexington, KY 40506, USA}

\begin{abstract}

Weak interaction contributions to hyperfine splitting and Lamb shift in light electronic and muonic atoms are calculated. We notice that correction to hyperfine splitting turns into zero for deuterium. Weak correction to the Lamb shift in hydrogen is additionally suppressed in comparison with other cases by a small factor $(1-4\sin^2\theta_W)$.

\end{abstract}


\preprint{UK/12-01}

\maketitle

\section{Introduction}

Unexpected results of the Lamb shift measurement in muonic hydrogen at PSI \cite{pohletal} gave rise to the proton radius puzzle. It consists of the five sigma discrepancy in the value of the proton radius extracted from the muonic hydrogen experiment \cite{pohletal} on the one hand, and the values extracted from the electronic hydrogen (see review in the CODATA compilation \cite{mtn2008}) and electron-proton scattering \cite{bern2010} on the other hand. A burst of theoretical activity followed the PSI experiment. Old results on the Lamb shift in muonic hydrogen were recalculated and confirmed (see, e.g., \cite{borie2011} and references therein), proton structure and polarizability corrections were critically reevaluated and improved (\cite{carlvan2011,hillpaz2011} and references therein), possible new physics explanations were explored (\cite{barger2011,batell2011} and references therein). Despite all these efforts no resolution of the proton radius puzzle was found. It seems now that solution of this problem will require a lot of additional experimental and theoretical work, in particular precise measurements of different transition frequencies in muonic hydrogen and other light muonic atoms. An experimental program on measurement of transition frequencies in light muonic atoms is now in progress by the CREMA collaboration at PSI. First experimental data on hyperfine splitting (HFS) in muonic hydrogen is coming soon, and the results for  muonic deuterium and helium ion are to follow \cite{kottm2011}. In anticipation of these experimental data we calculate below weak interaction contributions to HFS and Lamb shift in light muonic atoms, generalizing old results for muonic hydrogen \cite{eides1996} (see also \cite{esteve}).

Effective low-energy field-theoretic weak interaction Hamiltonian due to neutral currents for the fundamental fermions  has the form (see, e.g., \cite{naka2010})

\beq  \label{feremiint}
H_Z=\frac{4G_F}{\sqrt{2}}\int d^3x\left(\sum_i\bar\psi_i\gamma^\mu
(\widetilde T_3-\sin^2\theta_W Q)\psi_i\right)^2,
\eeq

\noindent
where $\theta_W$ is the Weinberg angle, $Q$ is the charge operator in terms of proton charge, $\widetilde T_3=T_3(1-\gamma^5)/2$, $T_3$ is the weak isospin, and summation goes over all species of fermions. Each current in this local four-fermion Hamiltonian contains a vector and an axial part.  For nucleons axial parts are renormalized by strong interactions and should be multiplied by $g_A=1.27$ (see, e.g., \cite{naka2010}). Specializing for the case of lepton-nucleon interaction the Hamiltonian in \eq {feremiint} reduces to (an extra factor two arises because all fields enter each factor in \eq{feremiint})

\beq \label{fourfff}
\begin{split}
H_Z&=\frac{G_F}{2\sqrt{2}}\int d^3x\left[\bar\psi_l \gamma^\mu\gamma^5\psi_l
-\bar\psi_l\gamma^\mu
\left(1-4\sin^2\theta_W \right)\psi_l\right]
\\
&\times\left[g_A\bar\psi_n\gamma_\mu\gamma^5\psi_n
-\bar\psi_n\gamma_\mu\psi_n-g_A\bar\psi_p\gamma_\mu\gamma^5\psi_p
+\bar\psi_p\gamma_\mu
\left(1-4\sin^2\theta_W\right)\psi_p\right],
\end{split}
\eeq

\noindent
where $\psi_l$ is the lepton (electron or muon) field, and $\psi_p$ and $\psi_n$ are the proton and neutron fields, respectively. This Hamiltonian generates all weak interaction contributions considered below.

\section{Weak Interaction Contributions to Hyperfine Splitting}

The leading weak interaction contribution to HFS arises from interaction of axial currents in \eq{fourfff}. In the leading nonrelativistic approximation only spatial components of axial neutral currents give nonzero contributions \cite{eides1996}, and the Hamiltonian simplifies

\beq
\begin{split}
H_Z&\to\frac{g_AG_F}{2\sqrt{2}}\int d^3x\left(\bar\psi_l \gamma^\mu\gamma^5\psi_l\right)
\left(\bar\psi_n\gamma_\mu\gamma^5\psi_n
-\bar\psi_p\gamma_\mu\gamma^5\psi_p\right)
\\
&\to-\frac{g_AG_F}{2\sqrt{2}}\int d^3x\left(\bar\psi_l \gamma^i\gamma^5\psi_l\right)
\left(\bar\psi_n\gamma^i\gamma^5\psi_n
-\bar\psi_p\gamma^i\gamma^5\psi_p\right).
\end{split}
\eeq

For a nucleus with $Z$ protons and $A-Z$ neutrons this field-theoretic Hamiltonian in the nonrelativistic limit reduces to the quantum mechanical Hamiltonian

\beq \label{weakham}
H_Z=\frac{g_AG_F}{2\sqrt{2}}\bm\sigma_l\cdot\left(\sum_p\bm\sigma_p
-\sum_n\bm\sigma_n\right)\delta^{(3)}(\bm r).
\eeq

\noindent
Matrix elements of this operator give the leading weak interaction contributions to hyperfine splitting that is nonzero only in $S$ states. The only remaining task is to calculate expectation value of the scalar product taking into account the nuclear wave function. We consider below in parallel light electronic and muonic atoms and ions, but numerical results are provided only for muonic systems.

\subsection{Hydrogen}

In the case of muonic (electronic) hydrogen there is only one term in the nuclear factor in \eq{weakham}, and we immediately obtain the leading weak interaction contribution to HFS splitting in the $nS$-state \cite{eides1996} in the form

\beq
\Delta E_Z(nS)=\frac{g_AG_F}{2\sqrt{2}}|\psi_n(0)|^3
(\bm\sigma_e\cdot\bm\sigma_p)|^{F=1}_{F=0},
\eeq

\noindent
where $\psi_n(0)=\sqrt{(Z\alpha m_r)^3/(\pi n^3)}$ is the Coulomb-Schr\"odinger wave function at the origin ($Z=1$ for hydrogen), $m_r=m_lm_p/(m_l+m_p)$ is the reduced mass, $\bm J=\bm\sigma_l/2$ is the lepton spin operator, $\bm I=\bm\sigma_p/2$ is the proton (nucleus) spin operator, $\bm F=\bm I+\bm J$ is the total angular momentum. Obviously, $(\bm\sigma_e\cdot\bm\sigma_p)|^{F=1}_{F=0}=4$ and

\beq
\Delta E_Z(nS)=
\frac{2g_AG_F}{\sqrt{2}}\frac{(Z\alpha m_r)^3}{\pi n^3}.
\eeq

\noindent
Numerically for $n=2$ in muonic hydrogen the weak contribution is

\beq
\Delta E_Z(2S)=2.8\times 10^{-4}~\mbox{meV},
\eeq

\noindent
what is at least an order of magnitude smaller than the uncertainty in hyperfine splitting due to  proton structure contributions \cite{cng2011}.

To elucidate the magnitude of the weak interaction contribution let us compare it with the dominant Fermi contribution to hyperfine splitting in muonic hydrogen is (see, e.g., \cite{egsreview,egsbook})

\beq
E_F=\frac{4}{3}g_p\frac{\alpha(Z\alpha)^3m_r^3}{m_lm_p}\approx 182.44~\mbox{meV},
\eeq

\noindent
where $g_p\approx 5.58\ldots$ \cite{mtn2008} is the proton $g$-factor in nuclear magnetons.

For excited states the dominant contribution to hyperfine splitting scales as $1/n^3$ and for the state with an arbitrary principal quantum number $n$ the ratio of the weak and dominant contributions to HFS in muonic hydrogen is

\beq
\frac{n^3\Delta E_Z(nS)}{E_F}
=\frac{3}{2\sqrt{2}\pi }\frac{g_AG_F m_\mu m_p}{g_p\alpha}\approx
1.2\ldots\times 10^{-5}.
\eeq

\subsection{Deuterium}

Deuteron is a spin one loosely bound system of two nonrelativistic nucleons that are predominantly described by the $S$-state wave function. Respective spin wave function is symmetric and the matrix element of the spin-symmetric deuteron nuclear factor of the effective Hamiltonian in \eq{weakham} in this approximation is equal zero,

\beq
\langle\bm\sigma_p -\bm\sigma_n\rangle=0.
\eeq

\noindent
This conclusion remains valid even with account of the admixture of the $D$ wave in the deuteron wave function, since the $D$ wave spin function is also symmetric with respect to spin variables (see, e.g., \cite{akhpom}). Hence, the weak interaction contribution to hyperfine splitting in electronic and muonic deuterium in the leading nonrelativistic approximation is zero.

\subsection{Tritium}

Triton is a spin one half ($I=1/2)$ system of one proton and two neutrons ($Z=1$, $A=3$). The third component of isospin for the triton is minus one half ($T_3=1/2-1/2-1/2=-1/2$). It is predominantly described by a product of the $S$ wave coordinate wave function and a completely antisymmetric spin-isospin wave function. Obviously, in this approximation   $\langle\bm\sigma_{p}-\bm\sigma_{n_1} -\bm\sigma_{n_2}\rangle=2\bm I$. A more accurate analysis with account for other components of the triton wave function produces  \cite{frpa2005}

\beq
\langle\bm\sigma_{p}-\bm\sigma_{n_1} -\bm\sigma_{n_2}\rangle=2\bm I\left(1-\frac{4}{3}P_{S'}-\frac{2}{3}P_D\right)=2c\bm I,
\eeq

\noindent
where $c\approx0.92$.

Further calculations go exactly like in the hydrogen case above and we obtain

\beq
\Delta E_Z(nS)=\frac{2cg_AG_F}{\sqrt{2}}\frac{(Z\alpha m_r)^3}{\pi n^3}.
\eeq

\noindent
Numerically for $n=2$ in muonic tritium the weak contribution is

\beq
\Delta E_Z(n=2)=3.1\times 10^{-4}~\mbox{meV}.
\eeq

Like in the hydrogen case we compare the weak contribution with the dominant Fermi energy in muonic tritium

\beq
E_F=\frac{4}{3}g_t\frac{\alpha(Z\alpha)^3m_r^3}{m_l m_p}
=239.919\ldots~\mbox{meV},
\eeq

\noindent
where $g_t=5.957 924 896(76)$ \cite{mtn2008} is the triton $g$-factor in nuclear magnetons.

For excited states the dominant contribution to hyperfine splitting scales as $1/n^3$, and for the state with an arbitrary principal quantum number $n$ the ratio of the weak and dominant contributions to HFS in muonic tritium is

\beq
\frac{n^3\Delta E_Z(nS)}{E_F}
=\frac{3}{2\sqrt{2}\pi}\frac{cg_AG_Fm_\mu m_p}{g_t\alpha}
\approx 1.0\ldots\times 10^{-5}.
\eeq

\subsection{Helium Ion}

Helion is a spin one half ($I=1/2$) system of two protons and a neutron ($Z=2$, $A=3$). The third component of isospin for the helion is one half ($T_3=1/2+1/2-1/2=1/2$). Like the triton the helion is predominantly described by a product of the $S$ wave coordinate wave function and a completely antisymmetric spin-isospin wave function. Obviously, in this approximation   $\langle\bm\sigma_{p_1}+\bm\sigma_{p_2} -\bm\sigma_{n}\rangle=-2\bm I$. A more accurate analysis with account for other components of the helion wave function produces \cite{frpa2005}

\beq
\langle\bm\sigma_{p_1}-\bm\sigma_{p_2} -\bm\sigma_{n}\rangle=-2\bm I\left(1-\frac{4}{3}P_{S'}-\frac{2}{3}P_D\right)=-2c\bm I.
\eeq

Further calculations go exactly like in the hydrogen and tritium cases above and we obtain

\beq
\Delta E_Z(nS)=-\frac{2cg_AG_F}{\sqrt{2}}\frac{(Z\alpha m_r)^3}{\pi n^3}.
\eeq

\noindent
Numerically for $n=2$ in muonic helium the weak contribution is

\beq
\Delta E_Z(n=2)=-2.5\ldots\times 10^{-3}~\mbox{meV}.
\eeq

Like in the hydrogen and tritium cases we compare the weak contribution with the dominant Fermi energy in muonic helium

\beq
E_F=\frac{4}{3}g_h\frac{\alpha(Z\alpha)^3m_r^3}{m_\mu m_p}=-1370.8\ldots~\mbox{meV},
\eeq

\noindent
where $g_h=-4.255 250 613$ \cite{mtn2008} is the helion $g$-factor in nuclear magnetons.

For excited states the dominant contribution to hyperfine splitting scales as $1/n^3$, and for the state with an arbitrary principal quantum number $n$ the ratio of the weak and dominant contributions to HFS in muonic helium  is

\beq
\frac{n^3\Delta E^Z}{E_F}
=-\frac{3}{2\sqrt{2}\pi}\frac{cg_AG_Fm_\mu m_p}{g_hZ\alpha}
\approx 1.5\ldots\times 10^{-5}.
\eeq

\subsection{Helium Ions $e^4He^+$, $\mu^4He^+$}

Spin of $\alpha$-particle is zero and there is no hyperfine structure in $e^4He^+$, $\mu^4He^+$ helium ions, and respectively no weak interaction contribution to hyperfine structure.

\section{Leading Weak Interaction Contribution to Lamb Shift}

The leading weak interaction contribution to the Lamb shift arises from interaction of vector currents in \eq{fourfff}. In the leading nonrelativistic approximation only time components give nonzero contributions \cite{eides1996}, and the interaction Hamiltonian simplifies

\beq
\begin{split}
H_Z&\to\frac{G_F}{2\sqrt{2}}\int d^3x\left(\bar\psi_l\gamma^\mu
\left(1-4\sin^2\theta_W \right)\psi_l\right)
\left(\bar\psi_n\gamma_\mu\psi_n-\bar\psi_p\gamma_\mu
\left(1-4\sin^2\theta_W\right)\psi_p\right)
\\
&\to\frac{G_F}{2\sqrt{2}}\int d^3x\left(\bar\psi_l\gamma^0
\left(1-4\sin^2\theta_W \right)\psi_l\right)
\left(\bar\psi_n\gamma_0\psi_n-\bar\psi_p\gamma_0
\left(1-4\sin^2\theta_W\right)\psi_p\right).
\end{split}
\eeq

\noindent
For a nucleus with $Z$ protons and $A-Z$ neutrons this field-theoretic Hamiltonian in the nonrelativistic limit  reduces to the quantum mechanical Hamiltonian

\beq
H_Z=\frac{G_F}{2\sqrt{2}}\left(1-4\sin^2\theta_W \right)
\left(A-Z-Z\left(1-4\sin^2\theta_W\right)\right)\delta^{(3)}(\bm r).
\eeq

\noindent
Matrix element of this operator gives the leading weak interaction contributions to the Lamb shift that is nonzero only in $S$ states. We obtain an explicit expression for the leading weak correction in all light atoms in the form

\beq \label{weaklmb}
\Delta E_Z(nS)=\frac{G_F}{2\sqrt{2}}\left(1-4\sin^2\theta_W \right)\left[(A-Z)-Z\left(1-4\sin^2\theta_W\right)\right]
\frac{(m_rZ\alpha)^3}{\pi n^3}.
\eeq

\noindent
For $A=Z=1$ this result was obtained in \cite{eides1996}. It is interesting to notice that in muonic hydrogen due to $A=Z=1$ the weak contribution to the Lamb shift is additionally suppressed by a small factor $1-4\sin^2\theta_W\approx0.08$. This suppression disappears for all other light muonic systems.

Let us compare weak contribution to the Lamb shift with the dominant contribution. The principal contribution to the Lamb shift in light muonic atoms is generated by the diagram with the electron vacuum polarization insertion in the Coulomb photon, and was calculated long time ago \cite{galpom} (see also reviews in \cite{blp,egsreview,egsbook})

\beq
\Delta E_{nl}=-\frac{8\alpha(Z\alpha)^2 m_r}{3\pi n^3}Q^{(1)}_{nl}(\beta),
\eeq

\noindent
where

\beq  \label{qnlanal}
Q_{nl}^{(1)}(\beta)\equiv
\int_0^\infty \rho d\rho\int_1^\infty d\zeta
f_{nl}^2\left(\frac{\rho}{n}\right)e^{-2\rho\zeta\beta}
\left(1+\frac{1}{2\zeta^2}\right)
\frac{\sqrt{\zeta^2-1}}{\zeta^2},
\eeq
\beq
f_{nl}\left(\frac{\rho}{n}\right)\equiv \sqrt{\frac{(n-l-1)!}
{n[(n+l)!]^3}}\left(\frac{2\rho}{n}\right)^le^{-\frac{\rho}{n}}
L^{2l+1}_{n-l-1}\left(\frac{2\rho}{n}\right),
\eeq

\noindent
$L^{2l+1}_{n-l-1}(x)$ is the associated Laguerre polynomial \cite{landlif} , and $\beta=m_e/(m_r Z\alpha)$.

\noindent
For the experimentally relevant interval $2P-2S$  we obtain

\beq
\Delta E(2P-2S)=\Delta E_{21}-\Delta E_{20}
=\frac{\alpha(Z\alpha)^2 m_r}{3\pi }(Q^{(1)}_{20}(\beta)-Q^{(1)}_{21}(\beta)),
\eeq

\noindent
and

\beq
\frac{\Delta E_{Z}(L,2S)}{\Delta E(2P-2S)}=
\frac{3G_Fm_r^2Z\left(1-4\sin^2\theta_W \right)\left[(A-Z)-Z\left(1-4\sin^2\theta_W\right)\right]}
{16\sqrt{2}(Q^{(1)}_{20}(\beta)-Q^{(1)}_{21}(\beta))}.
\eeq

\noindent
This expression demonstrates once again that due to the condition $A=Z=1$ the weak interaction contribution to the Lamb shift is additionally suppressed by an extra factor  $1-4\sin^2\theta_W\approx0.08$ in comparison with the weak interaction contribution in other light muonic systems. For muonic hydrogen $\beta\approx0.7$, $Q^{(1)}_{20}(\beta)=0.056$, $Q^{(1)}_{21}(\beta)=0.0037$, and we obtain

\beq
\frac{\Delta E_{Z}(L,n=2)}{\Delta E(2P-2S)}
\approx-1.7\times 10^{-9}.
\eeq

\noindent
We see that the weak correction to the Lamb shift in muonic hydrogen is orders of magnitude smaller than relative error of the Lamb shift measurement \cite{pohletal}. It is also much smaller than uncertainties of  the proton structure corrections \cite{carlvan2011}.

\section{Conclusions}

We calculated the leading weak contributions to HFS and Lamb shift in light muonic atoms and ions. The leading correction to HFS in deuterium is zero because the deuteron weak interaction Hamiltonian is antisymmetric with respect to nucleon spin variables while the deuteron spin wave function is symmetric. Corrections to Lamb shift in hydrogen are additionally suppressed by the small factor $(1-4\sin^2\theta_W)$. This happens because all other nuclei contain neutrons that weakly interact with leptons without this suppression factor. In all cases weak corrections are much smaller than current experimental and theoretical errors.

\begin{acknowledgments}

I am deeply grateful to Franz Kottmann for a question that triggered this short note. This work was supported by the NSF grant PHY-1066054.

\end{acknowledgments}

\end{document}